\documentclass[10pt,a4paper]{article}
\usepackage{amsmath,amssymb,latexsym}

\def\AFOUR{%
\setlength{\textheight}{8.5in}%
\setlength{\textwidth}{5.75in}%
\setlength{\topmargin}{-0.375in}%
\hoffset=-.5in%
\renewcommand{\baselinestretch}{1.17}%
\setlength{\parskip}{6pt plus 2pt}%
}

\AFOUR                                           

\expandafter\ifx\csname amssym.def\endcsname\relax \else\endinput\fi
\expandafter\edef\csname amssym.def\endcsname{%
       \catcode`\noexpand\@=\the\catcode`\@\space}
\catcode`\@=11
\def\undefine#1{\let#1\undefined}
\def\newsymbol#1#2#3#4#5{\let\next@\relax
 \ifnum#2=\@ne\let\next@\msafam@\else
 \ifnum#2=\tw@\let\next@\msbfam@\fi\fi
 \mathchardef#1="#3\next@#4#5}
\def\mathhexbox@#1#2#3{\relax
 \ifmmode\mathpalette{}{\m@th\mathchar"#1#2#3}%
 \else\leavevmode\hbox{$\m@th\mathchar"#1#2#3$}\fi}
\def\hexnumber@#1{\ifcase#1 0\or 1\or 2\or 3\or 4\or 5\or 6\or 7\or 8\or
 9\or A\or B\or C\or D\or E\or F\fi}

\font\tenmsa=msam10
\font\sevenmsa=msam7
\font\fivemsa=msam5
\newfam\msafam
\textfont\msafam=\tenmsa
\scriptfont\msafam=\sevenmsa
\scriptscriptfont\msafam=\fivemsa
\edef\msafam@{\hexnumber@\msafam}
\mathchardef\dabar@"0\msafam@39
\def\dashrightarrow{\mathrel{\dabar@\dabar@\mathchar"0\msafam@4B}}
\def\dashleftarrow{\mathrel{\mathchar"0\msafam@4C\dabar@\dabar@}}

\def\ulcorner{\delimiter"4\msafam@70\msafam@70 }
\def\urcorner{\delimiter"5\msafam@71\msafam@71 }
\def\llcorner{\delimiter"4\msafam@78\msafam@78 }
\def\lrcorner{\delimiter"5\msafam@79\msafam@79 }
\def\yen{{\mathhexbox@\msafam@55}}
\def\checkmark{{\mathhexbox@\msafam@58}}
\def\circledR{{\mathhexbox@\msafam@72}}
\def\maltese{{\mathhexbox@\msafam@7A}}
\def\circledS{{\mathhexbox@\msafam@73}}

\font\tenmsb=msbm10
\font\sevenmsb=msbm7
\font\fivemsb=msbm5
\newfam\msbfam
\textfont\msbfam=\tenmsb
\scriptfont\msbfam=\sevenmsb
\scriptscriptfont\msbfam=\fivemsb
\edef\msbfam@{\hexnumber@\msbfam}
\def\Bbb#1{{\fam\msbfam\relax#1}}
\def\widehat#1{\setbox\z@\hbox{$\m@th#1$}%
 \ifdim\wd\z@>\tw@ em\mathaccent"0\msbfam@5B{#1}%
 \else\mathaccent"0362{#1}\fi}
\def\widetilde#1{\setbox\z@\hbox{$\m@th#1$}%
 \ifdim\wd\z@>\tw@ em\mathaccent"0\msbfam@5D{#1}%
 \else\mathaccent"0365{#1}\fi}
\font\teneufm=eufm10
\font\seveneufm=eufm7
\font\fiveeufm=eufm5
\newfam\eufmfam
\textfont\eufmfam=\teneufm
\scriptfont\eufmfam=\seveneufm
\scriptscriptfont\eufmfam=\fiveeufm
\def\frak#1{{\fam\eufmfam\relax#1}}

\csname amssym.def\endcsname

\makeatletter
\def\section{\@startsection {section}{1}{\z@}{-3.5ex plus -1ex minus
 -.2ex}{2.3ex plus .2ex}{\large\sc}}
\def\subsection{\@startsection{subsection}{2}{\z@}{-3.25ex plus -1ex minus
 -.2ex}{1.5ex plus .2ex}{\normalsize\sc}}
\makeatother

\makeatletter
\@addtoreset{equation}{section}

\makeatother

\newcommand{\nc}{\newcommand}
\newcommand{\rnc}{\renewcommand}

\nc{\chap}[1]{{\clearpage}%
\begin{center}%
{\noindent\underline{\large\sc #1}}{\addcontentsline{toc}{section}{#1}}%
\end{center}%
{\vspace*{0.3cm}}}

\nc{\subs}[1]{{\vspace*{0.2cm}}%
{\noindent\underline{\small\sc
#1}}{\addcontentsline{toc}{subsubsection}{#1}}%
{\vspace*{0.2cm}}}

\nc{\be}{\begin{equation}}
\nc{\ee}{\end{equation}}
\nc{\bea}{\begin{eqnarray}}
\nc{\eea}{\end{eqnarray}}

\nc{\trac}[2]{{\textstyle\frac{#1}{#2}}}

\nc{\ex}[1]{\mbox{e}^{\,\textstyle#1}}

\nc{\CC}{\Bbb{C}}
\nc{\HH}{\Bbb{H}}
\nc{\PP}{\Bbb{P}}
\nc{\RR}{\Bbb{R}}
\nc{\ZZ}{\Bbb{Z}}
\nc{\II}{\Bbb{I}}
\nc{\EE}{\Bbb{E}}
\nc{\TT}{\Bbb{T}}
\nc{\DD}{\mathrm{I}\!\mathrm{D}}

\rnc{\d}{\delta}

\nc{\symx}{\circledS}
\newsymbol\smallsmile 1360
\newsymbol\smallfrown 1361
\nc{\ad}{\mathop{\mbox{ad}}\nolimits}
\nc{\tr}{\mathop{\mbox{tr}}\nolimits}
\nc{\Tr}{\mathop{\mbox{Tr}}\nolimits}
\nc{\Det}{\mathop{\mbox{Det}}\nolimits}
\rnc{\det}{\mathop{\mbox{det}}\nolimits}
\nc{\rk}{\mathop{\mbox{rk}}\nolimits}
\nc{\del}{\partial}
\nc{\diag}{\mathop{\mbox{diag}}\nolimits}
\nc{\ra}{\rightarrow}
\nc{\Ra}{\Rightarrow}
\nc{\LRa}{\Leftrightarrow}
\nc{\lra}{\leftrightarrow}
\nc{\ot}{\otimes}
\rnc{\ss}{\subset}
\nc{\nul}{\noindent\underline}
\nc{\non}{\nonumber\\}
\nc{\mat}[4]{\left(\begin{array}{cc}#1&#2\\#3&#4\end{array}\right)}
\rnc{\lg}{\frak{g}}
\nc{\G}[3]{\Gamma^{#1}_{\;{#2}{#3}}}
\nc{\nam}{\nabla_{\mu}}
\nc{\nan}{\nabla_{\nu}}
\nc{\dx}{\dot{x}}
\nc{\dxl}{\dot{x}^{\la}}
\nc{\dxm}{\dot{x}^{\mu}}
\nc{\dxn}{\dot{x}^{\nu}}
\nc{\ddx}{\ddot{x}}
\nc{\ddxm}{\ddot{x}^{\mu}}
\nc{\ddxn}{\ddot{x}^{\nu}}
\nc{\dxi}{\dot{\xi}}
\nc{\ddxi}{\ddot{\xi}}
\nc{\lsf}{\ell_s^{\mathrm{eff}}}
\nc{\lpf}{\ell_p^{\mathrm{eff}}}
\nc{\sqg}{\sqrt{g^{11}}}

\begin{document}


\vspace*{2cm}
\begin{center}
{\Large\sc Multiple M2-Branes and Plane Waves}
\end{center}
\vspace{0.2cm}

\begin{center}
{\large\sc Matthias Blau${}^a$} \textsc{and} 
{\large\sc Martin O'Loughlin${}^b$}\\[.8cm]
{\it ${}^a$ Institut de Physique, Universit\'e de Neuch\^atel, 
Breguet 1, Neuch\^atel, Switzerland}\\[.3cm]
{\it ${}^b$ University of Nova Gorica, Vipavska 13, 5000 Nova Gorica, Slovenia}
\end{center}

\vspace{1cm}

We propose a natural generalisation of the BLG multiple M2-brane
action to membranes in curved plane wave backgrounds, and verify in
two different ways that the action correctly captures the non-trivial
space-time geometry.  We show that the M2 to D2 reduction of the theory
along a non-trivial direction in field space is equivalent to the D2-brane
world-volume Yang-Mills theory with a non-trivial (null-time dependent)
dilaton in the corresponding IIA background geometry.  As another
consistency check of this proposal we show that the properties of metric
3-algebras ensure the equivalence of the Rosen coordinate version of this
action (time-dependent metric on the space of 3-algebra valued scalar
fields, no mass terms) and its Brinkmann counterpart (constant couplings
but time-dependent mass terms). We also establish an analogous result
for deformed Yang-Mills theories in any dimension which, in particular,
demonstrates the equivalence of the Rosen and Brinkmann forms of the
plane wave matrix string action.


\newpage

\section{Introduction}

The recent Bagger-Lambert-Gustavsson (BLG) proposal for a world-volume
theory of multiple membranes \cite{bl,gu}, following earlier work
\cite{preblg}, in terms of a 3-algebra gauge theory has already received
considerable attention. Various properties of the BLG theory have been
analysed e.g.\ in \cite{m2d2,jhs,dopoblg}, and a generalisation of the
BLG theory to Lorentzian 3-algebras associated to ordinary Lie algebras
has been proposed in \cite{lor3}.  The role of the 3-algebra structure
for 1-loop corrections to the BLG theory has been
discussed in \cite{gu2}. 

At the moment it is not completely clear \cite{bansen,dopolortz3} if the
Lorentzian 3-algebras really give a theory of multiple uncompactified
membranes in 11 dimensions or if they just provide an exotic rewriting
of the D2-brane world-volume theory \cite{d2d2,hisz}, and an alternative
generalisation of the BLG theory has been proposed in \cite{abjm}.

Nevertheless, deformations of the (generalised) BLG theories
\cite{bl,gu,lor3} may provide a Lagrangian description of multiple
M2-branes in non-trivial backgrounds and may also, in any case, be
of interest in their own right. Given the scarcity and rigidity of
finite-dimensional 
Euclidean \cite{beforelorentz3} and Lorentzian \cite{jose} 3-algebras,
one has to look elsewhere for suitable modifications. Certain 
mass \cite{massdefs} and Janus-like \cite{janusdefs}
deformations have already been considered, other variations of
the BLG action are discussed in \cite{other}, and a unified description
of various deformations of the BLG theory has been given in \cite{eric}.

In this same spirit, but along somewhat different lines,
we propose that the 3-algebra action with scalar sector
\begin{multline}
\label{isrcblg}
S_{RC-BLG}[X^I] = \int d^n\sigma\Tr\left(
-\trac{1}{2}g_{IJ}(t) (D_{\alpha} X^I, D^{\alpha} X^J) 
\right. \\ \left.
-\trac{1}{2.3!} g_{IL}(t)g_{JM}(t)g_{KN}(t)([X^I,X^J,X^K],[X^L,X^M,X^N]) 
\right)
\end{multline}
describes (for $n=3$)
multiple membranes 
extended along the $(x^\pm,x^9)$-directions
in the general Rosen coordinate (RC)
plane wave background
\be
\label{mpwrc}
ds^2 = 2 dx^+ dx^- + (dx^9)^2 + \sum_{I,J=1}^8 g_{IJ}(x^+) dx^I dx^J
\;\;,
\ee
in the same way that the BLG action 
(to which it reduces for $g_{IJ}(t) = \d_{IJ}$) describes
membranes in flat space 
(or some suitable M-orbifold thereof \cite{dopoblg}).\footnote{See
\cite{parks} for a complementary discussion of non-trivial backgrounds from
the M5-brane Nambu-Goto action point of view.}

In the absence of any straightforwardly applicable symmetry
considerations (the above Lagrangian will in general have 
no global symmetries, and the total action, with fermions, 
is not expected to have any linearly realised supersymmetries, 
since plane wave backgrounds are generically 1/2 BPS),
we will perform two other consistency checks on this proposal
which show that the action \eqref{isrcblg} correctly captures the
plane wave space-time geometry.

First (section 3)
we consider the analogue of the M2 to D2 \cite{m2d2} reduction
procedure for the Lorentzian 3-algebras \cite{lor3} (perhaps more
appropriately referred to as D2 to D2 \cite{d2d2}) in the presence
of a non-trivial metric component $g_{88}(x^+)$ along the
direction $X^8$ in field space that is being vev'ed.  We show 
that the resulting 2+1 dimensional Yang-Mills theory has
an effective time-dependent Yang-Mills coupling constant
\be
g^2_{YM}(t) = g_{88}(t) g^2_{YM} \;\;,
\ee
and that this is identical to the coupling constant one finds from
the world-volume theory of multiple D2-branes in the presence of a
non-trivial dilaton (with $x^8$ considered as the compactified
M-theory direction).

Another consistency check is provided by the
observation (section 4) that the 
action \eqref{isrcblg}
is related to the apparently completely different 3-algebra
action 
\begin{multline}
\label{isbcblg}
S_{BC-BLG}[Z^A] = \int d^n\sigma\Tr\left(
-\trac{1}{2}\d_{AB} (D_{\alpha} Z^A, D^{\alpha} Z^B) 
+\trac{1}{2}A_{AB}(t)(Z^A,Z^B) 
\right. \\ \left.
- \trac{1}{2.3!}\d_{AD}\d_{BE}\d_{CF} ([Z^A,Z^B,Z^C],[Z^D,Z^E,Z^F])
\right)
\end{multline}
(no time-dependent couplings on the scalar field space but arbitrary
time-dependent mass terms instead, encoded in the matrix $A_{AB}(t)$) 
by a simple linear transformation
of the fields, 
\be
\label{isrcsbc}
S_{RC-BLG}[X^I=E^I_{\;A}Z^A] = S_{BC-BLG}[Z^A]\;\;.
\ee
The validity of
\eqref{isrcsbc} provides strong evidence that \eqref{isrcblg}
and \eqref{isbcblg} encode the plane wave geometry \eqref{mpwrc}, 
since it should be regarded as the 3-algebra
field-theory counterpart of the statement
that a plane wave can also be written in the more common 
Brinkmann coordinates (BC) $z^{\mu}$ with, in particular,
$x^I=E^I_{\;A}z^A$ \eqref{rcbc2} as 
(suppressing the trivial $x^9$-direction)
\be
\label{imetrcbc}
2dx^+dx^- + g_{IJ}(x^+)dx^I dx^J =
 2dz^+dz^- + A_{AB}(z^+)z^Az^B (dz^+)^2 + \d_{AB}dz^A dz^B\;\;.
\ee
Note that in these coordinates, the membrane is stretched along the
metrically non-trivial $(z^\pm,z^9)$-directions. Thus the dependence
of the induced world-volume metric on the transverse coordinates via the
quadratic $A_{AB}z^Az^B$-terms manifests itself through mass terms in the
action \eqref{isbcblg}, as in the case of strings in the lightcone gauge.
This also provides us with a geometric interpretation of an arbitrary mass
deformation of the BLG theory in terms of plane waves (in the absence
of fluxes, we should also require the 11d vacuum Einstein equations to
be satisfied, namely that $A_{AB}$ be traceless). 

Note also that the BC form of the action \eqref{isbcblg} explains why
we focus on plane wave space-times here (since in principle we
could have e.g.\ allowed the couplings $g_{IJ}$ in \eqref{isrcblg} to
depend on all the world-volume coordinates). First of all, the analogy
with the quantisation of strings in the lightcone gauge suggests that
the BLG action is itself a lightcone gauge fixed action. Such a gauge
fixing is typically still 
possible e.g.\ for more general pp-wave backgrounds in which
the wave profile $A_{AB}(z^+,z^A)$ is not quadratic, but in that case
we would have to address the issue of how to define 
$\Tr(Z^A_{1},\ldots , Z^A_{k})$ for $k\neq 2$, while the quadratic mass term in \eqref{isbcblg}
is unambiguous. For the same reason we would also not want to
consider a dependence of $g_{IJ}$ in \eqref{mpwrc} or \eqref{isrcblg}
on the transverse coordinates $x^I$ or scalars $X^I$.

The 3-algebra gauge invariance of the actions, in particular the
existence of the invariant scalar product $\Tr(\;.\;)$, turns out to
play a crucial role in the proof of \eqref{isrcsbc}.  Along the way, we
will also establish an analogous result for Yang-Mills theories, which
implies in particular the equivalence of the Rosen and Brinkmann versions
of the matrix string action for plane waves \cite{mmpwbb}.\footnote{In
the spirit of the CSV matrix big bang model \cite{csv}, these provide a
non-perturbative description of string theory in a plane wave background -
see \cite{mmpwbb} for details and further references, since matrix string
theory is not our main concern in this short note.}

\section{Plane Wave Yang-Mills and 3-Algebra Actions}

The scalar sector of a prototypical non-Abelian Yang-Mills + scalar
action in $n$ space-time dimensions 
(with the flat world-volume metric $\eta_{\alpha\beta}$)
has the form 
\be
\label{sym}
S_{YM} = \int d^n\sigma\Tr\left(
-\trac{1}{2}\d_{IJ} D_{\alpha} X^I D^{\alpha} X^J 
- \trac{1}{4}g^2_{YM}\d_{IK}\d_{JL} [ X^I, X^J][X^K,X^L]\right)\;\;.
\ee
To bring out the analogies with, and differences to,
the 3-algebra actions, we recall here that
the $ X^I= X^I_a T^a$ are adjoint (Lie algebra valued) scalar fields,
$[T^a,T^b] = f^{ab}{}_{c} T^c$, 
$\Tr$ is an invariant scalar product (under the ad-action $[T^a,\;]$, which
acts as a derivation of the Lie bracket - the Jacobi identity)
on the Lie algebra, 
\be
\label{adinv1}
\Tr [T^a,T^b] T^c + \Tr T^b [T^a,T^c] = 0\;\;,
\ee
and the covariant derivative is
$D_{\alpha} X^I_a = \del_{\alpha} X^I_a - f^{bc}{}_a A_{\alpha\,b} X^I_c$.

Likewise the scalar sector of a 
prototypical 3-algebra action, namely the BLG action \cite{bl,gu}
(now blindly generalised to $n$ dimensions), is
\be
\label{sblg}
S_{BLG} = \int d^n\sigma\Tr\left(
-\trac{1}{2}\d_{IJ} (D_{\alpha} X^I, D^{\alpha} X^J) 
- \trac{1}{2.3!}\d_{IL}\d_{JM}\d_{KN} ([ X^I, X^J,X^K],[X^L,X^M,X^N])
\right).
\ee
Here the $ X^I= X^I_a T^a$ are 3-algebra valued scalar fields,
\be
\label{3rel}
{}[T^a,T^b,T^c] = f^{abc}{}_{d} T^d\;\;,
\ee
$\Tr (\;,\;)$ 
is an invariant scalar product (under the action of $[T^a,T^b,\;]$, which
acts as a derivation of the 3-bracket - the `fundamental identity')
on the 3-algebra, 
\be
\label{adinv2}
\Tr ([T^a,T^b,T^c], T^d) + \Tr (T^c, [T^a,T^b,T^d]) = 0\;\;,
\ee
and the covariant derivative is
$D_{\alpha} X^I_a = \del_{\alpha} X^I_a - f^{bcd}{}_a 
\mathcal{A}_{\alpha\,bc} X^I_d$.

These two basic classes of actions can be deformed in various ways,
e.g.\ by modifying the couplings of the scalar fields, and we will
consider two such modifications. 
The first class of actions arises from (\ref{sym}) or \eqref{sblg}
by replacing the flat
metric $\d_{IJ}$ on the scalar field space by a time-dependent matrix 
$g_{IJ}(t)$ of ``coupling constants''. Thus the deformed
action is (suppressing the coupling constant $g^2_{YM}$)
\be
\label{src}
S_{RC-YM} = \int d^n\sigma\Tr\left( 
-\trac{1}{2}g_{IJ}(t)D_{\alpha}X^I D^{\alpha}X^J
- \trac{1}{4} g_{IK}(t)g_{JL}(t)[X^I,X^J][X^K,X^L] 
\right)
\ee
in the Yang-Mills case, and \eqref{isrcblg} in the BLG case.
The second modification 
consists of simply adding (possibly time-dependent) mass terms for the scalars.
Thus, denoting the (same number of)
scalars in this model by $Z^A$, 
the actions we will consider are
\be
\label{sbc}
S_{BC-YM} = \int d^n\sigma\Tr\left( 
-\trac{1}{2}\d_{AB}D_{\alpha}Z^A D^{\alpha}Z^B
- \trac{1}{4} \d_{AC}\d_{BD}[Z^A,Z^B][Z^C,Z^D] +
\trac{1}{2}A_{AB}(t)Z^AZ^B \right)
\ee
and its 3-algebra counterpart \eqref{isbcblg},
with $A_{AB}(t)$ minus the mass-squared matrix. 

The labels RC and BC refer to the Rosen and Brinkmann coordinates of
plane wave metrics, as will become clear in section 4, and for this
reason we will also refer to the above actions and their 3-algebra
counterparts as plane wave actions.

\section{M2 (or D2) to D2 with a non-trivial Dilaton}

We consider the case where the metric \eqref{mpwrc} is of the form 
\be
\label{mpwrc2}
ds^2 = 2 dx^+ dx^- + (dx^9)^2 + \sum_{i,j=1}^7 g_{ij}(x^+) dx^i dx^j + 
g_{88}(x^+) (dx^8)^2\;\;,
\ee
and for the purposes of this section we may as well take
$g_{ij}=\d_{ij}$, since we just want to keep track of the effect of 
a non-trivial
$g_{88}$ in the M2 to D2 reduction \cite{m2d2,lor3,d2d2}.

We will also specifically consider the case of 
a Lorentzian 3-algebra \cite{lor3}, with generators 
$\{T^a\}= \{T^+,T^-,T^m\}$ and non-trivial structure constants
$f^{+mnp}=2f^{mnp}$, $f_{-mnp} = f_{mnp}$.
Expanding the 3-algebra valued fields 
in the above basis, $X^I = X^I_a T^a$, one finds that, 
as in \cite{lor3}, the field $X_-^I \equiv X^{+I}$
appears in the action \eqref{isrcblg} only via the term 
\be
\mathcal{L}_{X^+} 
= -\trac{1}{2} g_{IJ}(t)\del_{\alpha}X^{+I}\del^{\alpha} X^{-J}
+ \ldots,
\ee
leading to the equations of motion 
$\del_{\alpha} (g_{IJ}(t) \del^{\alpha} X^{-J})=0$.
A particular solution of this equation is $X^{-8}=\text{const}\neq 0$
and $X^{-i}=0$. Note, however, that there are other, non-constant,
solutions to this equation, even when $g_{IJ}(t)=\d_{IJ}$, employed
e.g.\ in \cite{janusdefs}, and that even for a constant solution here
we cannot appeal to $SO(8)$-invariance to rotate such a solution into
the $X^8$-direction. We are thus making the specific choice of
singling out this direction (corresponding to a specific M$\ra$IIA
reduction), and identify the vev of $X^{-8}$ with the Yang-Mills coupling
constant, $\langle X^{-8}\rangle = g_{YM}$.

It is now easy to see, by following the procedure in
\cite{lor3}, that the gauge invariant 
scalar kinetic term for $X^{8}_m$ and the
BF-term of the action,
\be
\begin{aligned}
\mathcal{L}_B &= -\trac{1}{2}g_{88}(t)
D_{\alpha} X^{8}_m
D^{\alpha} X^{8}_m+ 2 \epsilon^{\alpha\beta\gamma}B_{\alpha}^a
F_{\beta\gamma}^a + \ldots\\
D_\alpha X^{I}_m &= \del_{\alpha}X^{I}_m-2B_{\alpha\,m}X^{-I} +
f_{mnp}A^{n}_{\alpha}X^{pI}
\end{aligned}
\ee
combine to give rise to a Yang-Mills action 
\be
\mathcal{L}_{YM} =
-\frac{1}{4g^2_{YM}(t)} F^m_{\alpha\beta}F^{m\,\alpha\beta} +
\ldots
\ee
with the time-dependent coupling constant 
\be
\label{gymt}
g^2_{YM}(t) = g_{88}(t)g^2_{YM}\;\;.
\ee
This same combination also arises from the
sextic potential as the coefficient of the quartic potential term
for the remaining 7 scalar fields $X^i= X^{i}_m T^m$, and thus we can
indeed
identify it as the time-dependent coupling constant of the resulting
Yang-Mills theory.

How does this compare with the expectation that, somehow \cite{bansen},
this procedure of giving a vev to a scalar should correspond \cite{m2d2}
to compactifying M-theory on a circle down to IIA? Since the standard
relation
\be
ds^2 = \ex{-2\phi/3}ds_{st}^2 + \ex{4\phi/3}(dx^8)^2
\ee
between the M-theory and IIA string frame backgrounds implies that
$g_{88}=\exp 4\phi/3$, while the YM coupling constant of the D2-brane 
theory is usually set by $g^2_{YM} = g_s/\ell_s$, 
on the face of it this looks
incompatible with \eqref{gymt}. However, we have to remember that in
the string frame metric
\be
ds_{st}^2 = \ex{2\phi/3}(2 dx^+ dx^- + (dx^9)^2 + g_{ij}(x^+) dx^i dx^j) 
\ee
the induced metric $h_{\alpha\beta}$ 
on the D2-brane world-volume is non-trivial. Thus the 
D2-brane Yang-Mills action has the form
\be
\frac{1}{g_s\ell_s^3}\int d^3\sigma\; \ex{-\phi} \sqrt{-\det h}\;
h^{\alpha\gamma}h^{\beta\delta}\ell_s^4 \;
F^m_{\alpha\beta}F^m_{\gamma\delta}
=
\frac{\ell_s}{g_s}\int d^3\sigma\;\ex{-4\phi/3} 
\d^{\alpha\gamma}\d^{\beta\delta}
F^m_{\alpha\beta}F^m_{\gamma\delta}\;\;,
\ee
from which we read off the coupling constant
\be
g^2_{YM}(t) = (g_s/\ell_s)\ex{4\phi(t)/3}
\ee
(in the lightcone gauge $x^+=t$).
This agrees precisely with the 
result \eqref{gymt} obtained from `Higgsing' the Lorentizan BLG action.

\section{Rosen vs Brinkmann Form of Plane Wave Actions}

The purpose of this section is to establish that
the two, apparently rather different, classes of
Yang-Mills and 3-algebra 
actions introduced in section 2
are simply related by a linear, but time-dependent,
field redefinition $X^I=E^I_{\;A}(t)Z^A$  of the scalar fields, 
\be
\label{srcsbc}
S_{RC-YM/BLG}[X^I=E^I_{\;A}Z^A] = S_{BC-YM/BLG}[Z^A]\;\;.
\ee
As mentioned in the Introduction and explained in \cite{mmpwbb},
for the YM theories this claim originates from the equivalence of
the matrix string theory description of plane wave backgrounds in
Rosen and Brinkmann coordinates \eqref{imetrcbc}, and \eqref{srcsbc}
is the generalisation of this assertion to arbitary dimension $n$,
any number of scalar fields, and to 3-algebra actions.

We could straightaway prove \eqref{srcsbc} by a brute-force
calculation, but this would be rather unenlightening. Instead, we
will first consider a much simpler classical mechanics toy model
of this equivalence. We will then readily be able to establish the
result for the plane wave Yang-Mills actions (\ref{src},\ref{sbc}).
From this argument we then also learn how to use 3-algebra identities
to prove \eqref{srcsbc} in that case.

\subsection{A Classical Mechanics Toy Model}

Consider the Lagrangian 
$L_{bc}$ corresponding to the lightcone Hamiltonian 
of a particle in a plane wave in Brinkmann coordinates
(in the lightcone gauge $z^+=t$),
\be
L_{bc}(z) = \trac{1}{2}(\d_{AB}\dot{z}^A \dot{z}^B +A_{AB}(t)z^A z^B)\;\;,
\ee
and the corresponding Lagrangian in Rosen coordinates,
\be
L_{rc}(x) = \trac{1}{2}g_{IJ}(t)\dot{x}^I \dot{x}^J\;\;.
\ee
The claim is that these two Lagrangians are equal up to a total
time-derivative. To see this, recall that
the coordinate transformation 
between the Rosen and Brinkmann forms \eqref{imetrcbc} of a plane 
wave metric has the form
\be
\label{rcbc2}
(x^+,x^-,x^I) = (z^+ , 
z^- + \trac{1}{2}\dot{{E}}_{AI}{E}^I_{\;B}z^A z^B, 
{E}^I_{\;A} z^A)
\ee
where
$E^I_{\;A}=E^I_{\;A}(x^+) $ is a vielbein for $g_{IJ}(x^+)$ satisfying 
the symmetry condition 
\be
\dot{{E}}_{AI}{E}^I_{\;B} = \dot{{E}}_{BI}{E}^I_{\;A}\;\;,
\label{sc3}
\ee
and the relation between $g_{IJ}(x^+)$ and $A_{AB}(z^+)$ can be compactly 
written as \cite{mmhom}
\be
\label{aee}
A_{AB}(z^+)= \ddot{{E}}_{AI}(z^+) {E}^I_{\;B}(z^+)\;\;.
\ee
The symmetry condition \eqref{sc3}, which can be geometrically interpreted 
as the statement that the frame $E^I_{\;A}$ is parallel transported along
$\del_{x^+}$ \cite{bbop}, will play a crucial role on several
occasions in the following. 

Substituting $x^I=E^I_{\;A} z^A$ in $L_{rc}$, one can
now verify that one indeed
obtains $L_{bc}$ up to a total time-derivative. 
The way to see this without any calculation is to start with the complete
geodesic Lagrangian in Brinkmann or Rosen coordinates,
\be
\mathcal{L}= \trac{1}{2}g^{(rc)}_{\mu\nu}\dot{x}^{\mu}\dot{x}^\nu = 
\trac{1}{2}g^{(bc)}_{\mu\nu}\dot{z}^{\mu}\dot{z}^\nu 
\ee
in the lightcone gauge $z^+=x^+=t$, leading to
\be
L_{bc}(z) +  \dot{z}^- = L_{rc}(x) + \dot{x}^-\;\;.
\ee
This makes it manifest that the two Lagrangians $L_{bc}(z)$ and $L_{rc}(x)$
differ only by a total
time-derivative, namely the derivative of the shift of $x^-$ in the coordinate
transformation \eqref{rcbc2}.

A concrete illustration of this is provided by the standard harmonic
oscillator Lagrangian 
\be
\label{lbc}
L_{bc}(z) = \trac{1}{2}(\dot{z}^2 - \omega^2 z^2)\;\;,
\ee
whose equivalence with the somewhat more exotic Lagrangian
\be
\label{lrc}
L_{rc}(x) = \trac{1}{2}\sin^2 \omega t \;\dot{x}^2
\ee
with a time-dependent kinetic term can be traced back to the two
equivalent representations
\be
 2 dx^+dx^- + \sin^2 \omega x^+ (dx)^2
= 2dz^+dz^- - \omega^2 z^2 (dz^+)^2 + (dz)^2
\ee
of the corresponding plane wave geometry.

\subsection{Rosen to Brinkmann for Yang-Mills actions}

We can now come back to the two types of Yang-Mills actions
\eqref{src} and \eqref{sbc}, which are obviously in some
sense non-Abelian counterparts of the classical mechanics Brinkmann
and Rosen coordinate actions $S_{bc}=\int L_{bc}$ and
$S_{rc}=\int L_{rc}$ discuussed above. 
We are thus led to consider the linear transformation
\be
X^I(\sigma^\alpha) = E^I_{\;A}(t) Z^A(\sigma^\alpha)
\label{XEZ}
\ee
of the scalar fields, where
$E^I_{\;A}(t)$ is a vielbein for 
the metric (couplings) $g_{IJ}(t)$ on the scalar field space satisfying
(\ref{sc3}).

Even though in general non-Abelian coordinate transformations are
a tricky issue, this particular transformation is easy to deal with
since it is linear as well as 
diagonal in the Lie algebra. Consider
e.g.\ the quartic potential terms in (\ref{src}) and (\ref{sbc}). With
the substitution \eqref{XEZ}, one obviously has
\be
\label{pot}
\begin{aligned}
g_{IK}g_{JL} [X^I,X^J][X^K,X^L]&=
g_{IK}g_{JL} E^I_{\;A}E^J_{\;B}E^K_{\;C}E^L_{\;D} [Z^A,Z^B][Z^C,Z^D]\\
&= \d_{AC}\d_{BD} [Z^A,Z^B][Z^C,Z^D]\;\;,
\end{aligned}
\ee
so that the two quartic terms are indeed directly related by \eqref{XEZ}.
Now consider the gauge covariant kinetic term for the scalars in \eqref{src}.
Since $E^I_{\; A} = E^I_{\;A}(t)$ depends only on $t$, 
the spatial covariant derivatives transform as 
\be
\label{dnt}
\alpha \neq t:\quad D_{\alpha}X^I = E^I_{\; A}(t) D_{\alpha}Z^A\;\;,
\ee
so that the spatial derivative parts of the scalar kinetic terms are mapped
into each other. It thus remains to discuss the term 
$\Tr g_{IJ}(t)D_t X^I D_t X^J$
involving the covariant time-derivatives.
The term with two gauge fields $A$ is purely algebraic and is thus
mapped directly to its BC counterpart in the term $\Tr \d_{AB} D_t Z^A
D_t Z^B$.
For the term quadratic in the 
ordinary $t$-derivatives, the argument is identical to 
that in section 2.2, 
and thus, using \eqref{sc3} and \eqref{aee}, one finds 
\be
\label{kin}
\trac{1}{2}\Tr g_{IJ}(t) \dot{X}^I \dot{X}^J =
\trac{1}{2}\Tr(\d_{AB}\dot{Z}^A \dot{Z}^B +A_{AB}(t)Z^A Z^B) +
\trac{d}{dt}(\ldots)
\;\;.
\ee
The only remaining subtlety are terms involving the $t$-derivative 
$\dot{E}^I_{\;A}$ of $E^I_{\; A}$, arising from cross-terms like
\be
\label{mixed1}
\Tr g_{IJ}(t)[A_t,X^I] \del_t X^J = 
\Tr g_{IJ}(t)E^I_{\;A}[A_t,Z^A] \del_t (E^J_{\;B} Z^B) \;\;.
\ee
However, these terms do not contribute at all since
\be
\label{mixed2}
g_{IJ}(t)E^I_{\;A} \dot{E}^J_{\;B} \Tr [A_t,Z^A] Z^B
= g_{IJ}(t)E^I_{\;A} \dot{E}^J_{\;B} \Tr A_t[Z^A, Z^B] =0
\ee
by the ad-invariance of the trace \eqref{adinv1} and the 
symmetry condition \eqref{sc3}. This establishes \eqref{srcsbc}.

It is pleasing to see that this symmetry condition, which already
ensured several cancellations in the standard tranformation from
Rosen to Brinkmann cooordinates (and thus also in establishing e.g.\
\eqref{kin}), is also responsible for the elimination of some terms of
genuinely non-Abelian origin (something the symmetry condition was not
originally designed for).

The above equivalence is also valid for models with a time-dependent
dilaton/coupling constant, as in \cite{mmpwbb}, since the total
time-derivative arises only from the (dilaton-independent) scalar
kinetic term. In \cite{mmpwbb}, we illustrate the advantages of the
BC representation (the scalar kinetic term has the canonical form and
the mass term encodes invariant geometric information about the plane
wave since $A_{AB}(z^+)$ is its curvature tensor) \textit{vis-\`a-vis}
its RC counterpart in the matrix string context.

This equivalence also extends in a rather obvious way to the appropriate
fermionic terms of the action. In these models the couplings between
the fermions $\Psi$ and the scalar fields $X^I$ universally have the form
\be
S_{\Psi} \sim \int d^n\sigma\Tr \bar{\Psi} \Gamma_I[X^I,\Psi]\;\;.
\ee
Thus the only effect of the transformation \eqref{XEZ} is to convert the
RC gamma-matrices $\Gamma_{I}$ to their BC (frame component) counterparts
$\Gamma_A = E_A^I\Gamma_I$,
\be
\Gamma_I X^I = \Gamma_A Z^A\;\;,
\ee
with
\be
\{\Gamma_I,\Gamma_K\} = 2 g_{IK} \quad\Ra\quad
\{\Gamma_A,\Gamma_B\} = 2 \d_{AB} \;\;.
\ee

\subsection{Rosen to Brinkmann for 3-algebra actions}

We now consider the effect of the transformation 
$X^I= E^I_{\;A}(t) Z^A$ of the 3-algebra valued fields on 
the RC action \eqref{isrcblg}. It is straightforward to see that, 
exactly as in \eqref{pot}, the sextic potential term is mapped to that of 
the BC action \eqref{isbcblg}, and that the YM-theory identities 
\eqref{dnt} and \eqref{kin} remain valid in the 3-algebra context, 
so that in particular 
the mass terms of \eqref{isbcblg} are generated in this way.
It thus only remains to discuss, similarly to \eqref{mixed1}, the cross-terms
between a $t$-derivative and the 3-algebra connection. These have the form
\be
\Tr(T^a,T^c)\;(\partial_t X_a^I)
f^{deb}_{\phantom{adb}c} \mathcal{A}_{t\,de}
X_b^J g_{IJ}(t)
= \Tr(T^a,[T^d,T^e,T^b])\;
(\partial_t X_a^I) \mathcal{A}_{t\,de} X_b^J g_{IJ}(t)\;\;,
\ee
where we used the 3-algebra relation \eqref{3rel} in the form
$f^{deb}_{\phantom{edb}c}T^c = [T^d,T^e,T^b]$. 
Inserting the field transformation, we now find that there is
just one troublesome term, namely the one
involving the $t$-derivative of the
transformation matrix $E^I_{\;A}(t)$ itself. However, here again the symmetry
condition \eqref{sc3} and the invariance of the trace \eqref{adinv2} come to
the rescue to show that this term is identically zero, 
\be
\Tr(T^a,[T^d,T^e,T^b])\;
\dot{E}^I_{\;A} Z_a^A \mathcal{A}_{t\,de} E^{J}_{\;B} Z_b^B g_{IJ}(t) = 0
\;\;,
\ee
since $\Tr(T^a,[T^d,T^e,T^b])$ is anti-symmetric in the indices $a,b$
while the second part of the above expression is symmetric.

As in the YM case, the fermionic terms are also mapped to each other, since
the coupling between the fermions $\Psi$ (11d Majorana spinors subject to the
constraint $\Gamma_{012}\Psi=-\Psi$) and the scalar fields is purely algebraic
\cite{bl,gu,lor3}, so that
\be
\Tr(\bar{\Psi},\Gamma_{IJ}[X^I,X^J,\Psi]) = 
\Tr(\bar{\Psi},\Gamma_{AB}[Z^A,Z^B,\Psi]) \;\;.
\ee

The results of this and the previous section provide us with reasonable (albeit
still rather circumstantial) evidence that the deformed BLG actions
(\ref{isrcblg},\ref{isbcblg}) that we have proposed indeed describe
multiple M2-branes in a curved plane wave background, but much remains
to be understood regarding the BLG actions, their extension to curved
space-times, and their relation to multiple M2-branes in general.

\subsection*{Acknowledgements}

We are grateful to Denis Frank, Giuseppe Milanesi and Sebastian
Weiss for discussions and for their collaboration on related matters.
This work has been supported by the Swiss National Science Foundation
and by the EU under contract MRTN-CT-2004-005104.

\rnc{\Large}{\normalsize}

\end{document}